\documentclass[preprint,proceedings]{rmaa}

\usepackage{paralist}
\usepackage{psfrag,color}

\SetYear{2006}
\SetConfTitle{11th Latin-American Regional IAU Meeting}

\title{Starbursts and AGN Fueling through Secular Evolution } 

\author{ F. Combes,\altaffilmark{1} }

\altaffiltext{1}{LERMA, Observatoire de Paris}

\shortauthor{F. Combes}
\shorttitle{Secular Evolution}

\fulladdresses{
\item LERMA, Observatoire de Paris, 61 Av. de l'Observatoire, F-75014, Paris, France
  (\email{francoise.combes@obspm.fr}).
}

\listofauthors{F. Combes}
\indexauthor{Combes, F.}

\abstract{
Except in the most extreme cases of nuclear activity, either starbursts or AGN,
it is difficult to find observationnally a close link between the dynamics
and the activity. Theoretically however, the necessary step to fuel the
gas to the center, is that gravity torques are created through a non-axisymmetric
pattern, either bar and/or spiral, triggered or not by a tidal interaction.
  We describe the sequence of processes for a typical evolution
cycle for a spiral galaxy, and the possible efficient feedback mechanisms. 
The various morphologies and dynamical states of spiral galaxies are
 interpreted in terms of a sequence of evolutionary phases, and
the corresponding time-scales can be estimated from observations.
In this scenario, activity in galaxies is related to the appearance of bar instability, 
although they might not be synchronised in phase. The role of external gas accretion
in the secular evolution is discussed.
 }

\addkeyword{Galaxies: AGN}
\addkeyword{Galaxies: Dynamics}
\addkeyword{Galaxies: Starburst}
\addkeyword{Secular evolution}

\begin{document}
\maketitle

\section{Secular evolution versus violent interactions}

Galaxies may accumulate mass across the Hubble time
through essentially two ways: a violent way, accreting
satellites, interacting and merging with companions, and a 
slow and more continuous way, that we call secular evolution,
where the accretion is only from matter in the cosmic filaments,
helped by internal instabilities.

The role of internal dynamical processes, related to non-axisymmetric
perturbations such as bars or spirals, must be important
in the evolution of galaxies, since too many interactions
heat and destroy disks (e.g. Toth \& Ostriker 1992), 
while the majority of galaxies today are disk galaxies.

Secular Evolution (SE) may explain the 
formation and re-formation of spirals and bars,
and also may explain the formation of bulges, at
least the smaller ones. Among the spheroids, 
there is a large range of masses and dynamical properties,
they possess more or less rotation (in terms of $V_{rot}/\sigma$)
and a radial distribution obeying Sersic laws, spanning
all the range from exponential to de Vaucouleurs profiles.
Small bulges with nearly exponential profiles
and some rotation are called pseudo-bulges 
(see the review by Kormendy \& Kennicutt 2004).
Their kinematics show that they have been formed through
the disk (Bureau \& Athanassoula 2005, Gadotti \& de Souza 2005,
Debattista et al 2004).
Their intermediate properties also involve flattening and age/metallicity.
The importance of SE is revealed by the observation 
of the bulge-disk correlation in characteristic radii:
in average, their ratio is r$_e$/h = 0.22, and varies from 0.20 to 0.24 from
late to early types (MacArthur, Courteau \& Holtzman 2003).

In what follows, we present examples of dynamical phenomena
implying secular evolution. We show in particular
how the observed asymmetrical morphology of galaxies 
can help to quantify the importance of SE and of external
gas accretion in galaxy evolution.  When galaxies look peculiar, it is tempting
to attribute this to galaxy-galaxy interaction; but external mass accretion can also produce
disturbed morphologies. 

Secular evolution might also trigger starbursts: gas is not accreted
continuously but through the action of bars and resonances, the gas is 
infalling by intermittence towards the center and can lead to
nuclear bursts. These bursts can then be
followed by a cycle of nuclear activity. 

\section{Gas accretion to maintain SFR and bars}

It has been observed for a long time that
galaxies in the middle of the Hubble sequence have maintained 
their star formation rate about
constant over a Hubble time (Kennicutt 1983, Kennicutt et al 1994).
Even taking into account the stellar mass loss, to replenish
the interstellar medium, an isolated 
galaxy should have an exponentially decreasing star formation rate.
Part of the gas could come from accreted gas-rich dwarfs,
but this is far from sufficient, given the average gas mass
contained in these small systems. On the other hand,
major mergers with too massive systems are destructive
for  disks.

Another constraint on external gas accretion comes from
the presence of bars in spiral galaxies. Bars are a gravitational
instability that spontaneously develop in a cold disk of stars
and gas. However, bars appear to be self-destroying in the presence of gas,
when the latter corresponds to at least a few percent of the disk mass.
  This self-regulating process is part of secular evolution,
involving gravity torques exerted between the bar and the disk gas.
Bars are omnipresent in spiral galaxies today,
and this is not expected if the gas contained in
the galaxy disks inflows towards the center and destroys the bars.
A solution is that external gas is accreted at a sufficient rate in order
to replenish spiral disks, and reform a bar after
the previous one has vanished. 

Bar destruction has been discovered in numerical 
simulations more than a decade ago (e.g. Hasan \& Norman 1990, 
Friedli \& Benz 1993), but the mechanism thought 
to be responsible for the bar destruction was the existence
of a central mass concentration (CMC), driven by
the bar torques and the gas inflow (Hasan et al 1993).
  The apparition of the CMC changed the mass distribution in the center,
destroyed the structure of orbits supporting the bar, and increased
the extension of chaotic orbits.

\subsection{Angular Momentum Transfer}
Bars are waves with negative angular momentum (AM): 
they are created by a transfer of AM to the outer disk (through a transient spiral).
In collisionless systems (no gas), the exchanges can only be with the
outer stellar disk, when the dark matter halo is not
dominating the dynamics inside the baryonic disk.
In a halo-dominated system, with an NFW profile, the
angular momentum can be transfered to the dark matter,
although with a lower efficiency (Athanassoula 2002, Curir et al  2005).
 Constraints can then be put on the dark matter fraction
inside the bar radius, since the bar pattern speed can
be significantly decreased by dynamical friction of
the bar on the dark matter particles (Debattista \& Sellwood 2000).

When the dark halo is not dominant in the inner galaxy,
angular momentum transfer is very efficient with gas.
  The gas inflow is not essentially due to viscous torques,
which are quite weak over galactic scales (Lin \& Pringle 1987),
but to gravity torques.  The measurements of
the bar torques on the gas, from the near-infrared
images as mass tracers, and the observed
phase shift between the potential and gas surface density,
confirm the strength of the torques and the time-scale
for AM exchange (e.g. Garcia-Burillo et al 2005).

\subsection{Gas inflow and bar destruction}

The bar destruction rate was recently questioned, in particular by Shen \& Sellwood (2004)
who simulated the action of various CMC (different masses, different concentrations)
on the bar strength, and concluded that bars were more robust than previously thought.
But this just demonstrated that the CMC was not the main mechanism to destroy
bars in previous simulations, and that the gas inflow itself has to be
taken into account. Since the gas is driven in by the bar torques,
it exerts by reaction a torque on the bar of opposite sign, which
gives AM to the bar: the AM lost by the gas which flows towards 
the center is accepted by the bar, which then weakens.
The relative role of gas inflow and CMC is then clarified 
(Bournaud et al 2005b). The fact that the CMC is not
sufficient to destroy the bar explains why a bar can be reformed
easily, through external gas accretion (Bournaud \& Combes 2002).

The details of the gas inflow reveal a discontinuous
process. Gravity torques are negative only inside bar corotation (CR), and 
are positive ouside: therefore only the gas inside the bar region can be driven inwards,
when the bar is strong. The outer gas accumulates in a ring 
at the outer Lindblad resonance (OLR) (e.g. Buta \& Combes 1996). 
 Also, the external gas accreted by the galaxy potential well from
the cosmic filaments accumulate in the outer parts, and is stalled at OLR.
 This gas has to wait the bar destruction to be able to flow inwards,
due to viscous torques (these must be understood
as coming from macro-turbulence in the interstellar medium). 
The time-scale for the gas to flow in is then shortened with respect
to the continuous disk configuration, due to the ring distribution
(time-scale inversely proportional to the gradient of gas surface density). 

The global result is that external gas is accreted by intermittence,
with a typical time-scale equal to that of the bar formation
and destruction (of the order of 1 Gyr or less).
The bar cycle may be as short as a few 10$^8$ yrs, depending
on the bar strength, and on this time scale, the
gas inflow towards the center may produce nuclear starbursts,
and trigger an AGN cycle.
 This last step may require the presence of 
two inner Linblad resonances, and the decoupling
of a nuclear bar, without which the gas is stalled at the ILR.
Inflow with 2 embedded bars can then occur on very short time-scales,
of a few 10$^7$ yrs or less, and these largely different time-scales
explain why it is difficult to detect in the observations any clear correlation
between the presence of a large-scale bar and AGN activity in spiral galaxies
(Garcia-Burillo et al 2005).

The detailed distribution of bar strengths in spiral 
galaxies, compared with numerical predictions
can help to  quantify  the required the external gas accretion rate;
a typical galaxy has to double its mass in 10 Gyr though this mechanism
(Block et al 2002, Laurikainen et al 2004).
The fact that the first census at high redshift indicates
a constant bar fraction over several Gyr (Jogee et al 2004)
supports this conclusion.

\subsection{Feedbacks}

Secular evolution involves self-regulated cycles and feeback mechanisms such as bar self-destruction.
Other feedback have been invoked to play a large role in galaxy evolution, in moderating gas accretion;
some are related to star formation feedback (Thacker \& Couchman 2001), and others
to the AGN energy feedback, since a black hole is known
to exist in almost every galaxy (Di Matteo et al 2005). 
The duty cyle of an AGN phase is estimated to about 100 Myr,
the energy released by  the AGN quenches both star formation  and AGN growth,
in very massive galaxies (Croton et al 2006).

The best example of AGN feedback is provided by the moderation of cooling flows
in galaxy clusters, where recent X-ray observations have emphasized the
reheating processes of the radio jets through  schocks,  
acoustic waves, bubbles (e.g. Fabian et al 2003).
This kind of feedback is not likely to play a significant role
in the secular evolution of LLAGN, concerned with  mild evolution.

\section{Lopsided galaxies}

A consequence of external gas accretion may also be seen in 
the frequency of asymmetric and off-centered galaxies. This
affects in particular the extended disks of neutral gas around galaxies.

Lopsided galaxies have been revealed  by their asymmetric HI global spectrum
with a frequency of 50\% in the survey by Richter \& Sancisi (1994) of 1700 galaxies.
This frequency is even larger (77\%) in late-type galaxies (Matthews et al 1998).

Stellar disks are also observed lopsided (Zaritsky \& Rix 1997).
In a recent study of the near-infrared images
of the OSU sample (Eskridge et al 2002), the Fourier parameter  A1 ($m=1$
component power of the density, normalised to the $m=0$ one)
has been found larger than 0.2 in about 20\% of galaxies (Bournaud et al 2005a).
 Given the maximum asymmetry allowed by internal instabilities,
at least 2/3 of these galaxies have an $m=1$ perturbation required by
an external mechanism.

Can this be attributed to companion interactions?
It is not likely, since most lopsided galaxies  are isolated (Wilcots \& Prescott 2004).
The parameter A1 in the NIR OSU sample was found uncorrelated with the
tidal index $T_p \propto  (M/m)\,  (r^3/D^3$), with $M/m$ the mass ratio with companions,
and $r^3/D^3$ the cube of the size to distance ratio.
Furthermore, galaxy interactions cannot explain that A1 is higher in late-type
galaxies.

Simulations of minor mergers can produce an $m=1$ perturbation, but over a limited
time-scale. Only continuous and asymmetric gas accretion (with a few M$_\odot$/yr)
can explain the observed frequency of $m=1$
and the long life-time of the perturbation (Bournaud et al 2005a).

\section{Role of galaxy mergers}

 Of course, galaxy minor mergers also play a role in parallel
to secular evolution, and it is important to quantify their
effects in the morphology and dynamics of remnants,
to understand their relative past frequency, as a function of
the present state of spiral galaxies.
It is now well established that major mergers, with at least
a mass ratio equal to 1/3, can significantly destroy disks in
most geometries, and the remnants evolve towards ellipticals
(e.g. Naab \& Burkert 2003). Also, towards the very minor mergers side,
with mass ratios between 1/100 and 1/10, some works have 
quantified the degree of heating of the disk, that are
not destroyed but remain spiral disks (Walker et al 1996).

Remnants from intermediate-mass mergers have been 
studied, with mass ratios between 1/10 and 1/3
(Bournaud, Jog, Combes 2005). They can be characterized by hybrid systems,
surprisingly  with density profiles that can still be described by
exponentials over most radii, but with thicknesses more
reminding of lenticular systems.  However, the kinematics
is more similar to elliptical systems, in the sense that the amount
of rotation disappears rather quickly (in terms of $V_{rot}/\sigma$).
As can be expected, $V_{rot}/\sigma$  decreases regularly with the mass ratio, with
some modulation with the prograde/retrograde character
of the merger. 

With a small amount of successive minor mergers (3 for a mass ratio
of 1/7, or 5 for a mass ratio of 1/10), the remnant becomes a
genuine elliptical. It is likely that this mode of formation is more
frequent than a true major merger, given the mass function of galaxies.
This reveals again the
importance of gas accretion all along the evolution, to avoid
the formation of too many spheroids, without replenishing disks.

\section{Conclusion}

Numerical simulations, supported by strong constraints from observations,
reveal that secular evolution plays a fundamental role in the mass
accretion by galaxies, in particular in the bulge formation, and
in explaining the fueling of nuclear starbursts and AGN. 

The mass accretion cannot be mainly due to 
mergers with companions, since they are quite destructive
for disks. This requires more diffuse gas accretion from cosmic filaments,
which can be more continuous and soft, without stellar condensations.

The observed frequency of non-axisymmetries such as
bars and m=1 perturbations constrain the accretion rate,
such that a typical galaxy must double its mass in about 10 Gyr.

Dynamical feedback mechanisms self-regulate the cycles,
such as bar destruction through radial gas inflow, due to tranfer
of angular momentum. Other feedback may moderate gas accretion
such as star formation, and the energy of the AGN in very
massive systems.

Mergers have also their role in building spheroids and heating disks,
in particular mergers with intermediate mass ratios lead to 
hot and thick hybrid systems, that progressively lose their rotation.
 However, the relative fraction of secular versus merger
evolution is decreasing along the Hubble sequence.

\acknowledgements

I am very grateful to the organisers for their kind
\adjustfinalcols
invitation to this 11th LARIM in Pucon, Chile.

\end{document}